\documentclass[reprint,superscriptaddress,amsmath,amssymb,aps,prb,showpacs]{revtex4-1}
\usepackage{}

\usepackage{graphicx}
\usepackage{dcolumn}
\usepackage{bm}

\usepackage{natbib}
\usepackage[colorlinks=true,citecolor=blue,linkcolor=blue,pdfhighlight =/O]{hyperref}
\hypersetup{urlcolor=blue}
\usepackage{booktabs}

\newcommand{\pcod}{Pr$_{2}$CuO$_{4\pm\delta}$\,}
\newcommand{\pcco}{Pr$_{2-x}$Ce$_{x}$CuO$_{4}$\,}
\newcommand{\rcco}{R$_{2-x}$Ce$_{x}$CuO$_{4}$\,}
\newcommand{\lsco}{La$_{2-x}$Sr$_{x}$CuO$_{4}$\,}
\newcommand{\plcco}{Pr$_{1-x}$LaCe$_{x}$CuO$_{4}$\,}

\newcommand{\etal}{{\it et~al. }}

\begin{document}

\title{Tunable superconductivity in parent cuprate \pcod thin films}

\author{Xinjian Wei}
\affiliation{Beijing National Laboratory for Condensed Matter Physics, Institute of Physics, Chinese Academy of Sciences, Beijing 100190, China}
\affiliation{School of Physical Sciences, University of Chinese Academy of Sciences, Beijing 100049, China}

\author{Ge He}
\affiliation{Beijing National Laboratory for Condensed Matter Physics, Institute of Physics, Chinese Academy of Sciences, Beijing 100190, China}
\affiliation{School of Physical Sciences, University of Chinese Academy of Sciences, Beijing 100049, China}

\author{Wei Hu}
\affiliation{Beijing National Laboratory for Condensed Matter Physics, Institute of Physics, Chinese Academy of Sciences, Beijing 100190, China}
\affiliation{School of Physical Sciences, University of Chinese Academy of Sciences, Beijing 100049, China}

\author{Xu Zhang}
\affiliation{Beijing National Laboratory for Condensed Matter Physics, Institute of Physics, Chinese Academy of Sciences, Beijing 100190, China}
\affiliation{School of Physical Sciences, University of Chinese Academy of Sciences, Beijing 100049, China}

\author{Mingyang Qin}
\affiliation{Beijing National Laboratory for Condensed Matter Physics, Institute of Physics, Chinese Academy of Sciences, Beijing 100190, China}
\affiliation{School of Physical Sciences, University of Chinese Academy of Sciences, Beijing 100049, China}

\author{Jie Yuan}
\affiliation{Beijing National Laboratory for Condensed Matter Physics, Institute of Physics, Chinese Academy of Sciences, Beijing 100190, China}

\author{Beiyi Zhu}
\affiliation{Beijing National Laboratory for Condensed Matter Physics, Institute of Physics, Chinese Academy of Sciences, Beijing 100190, China}

\author{Yuan Lin}
\affiliation{State Key Laboratory of Electronic Thin Films and Integrated Devices $\&$ Center for Information in Medicine, University of Electronic Science and Technology of China, Chengdu 610054, China}

\author{Kui Jin}\thanks{kuijin@iphy.ac.cn}
\affiliation{Beijing National Laboratory for Condensed Matter Physics, Institute of Physics, Chinese Academy of Sciences, Beijing 100190, China}
\affiliation{School of Physical Sciences, University of Chinese Academy of Sciences, Beijing 100049, China}
\affiliation{Collaborative Innovation Center of Quantum Matter, Beijing 100190, China}

\date{\today}

\begin{abstract}
In this article, we studied the role of oxygen in \pcod thin films fabricated by polymer assisted deposition method. The magnetoresistance and Hall resistivity of \pcod samples were systematically investigated. It is found that with decreasing the oxygen content, the low-temperature Hall coefficient ($R_H$) and magnetoresistance change from negative to positive, similar to those with the increase of Ce-doped concentration in \rcco (R= La, Nd, Pr, Sm, Eu). In addition, $T_c$ versus $R_H$ for both \pcco and \pcod samples can coincide with each other. We conclude that the doped electrons induced by the oxygen removal are responsible for the superconductivity of $T^\prime$-phase parent compounds.
\end{abstract}

\pacs{ 74.72.Ek,   81.15.-z,  73.50.-h, }

\maketitle

\section{INTRODUCTION}

Studying parent compounds of high-$T_c$ superconductors is crucial to unveiling the mechanism of high-temperature superconductivity, as well as provides significant clues to explore new superconductors. Parent compounds of the cuprates have been long considered as antiferromagnetic Mott insulators, which can be tuned to superconductors by doping holes or electrons \cite{Lee2006}. For example, Sr$^{2+}$ cations substitutes La$^{3+}$ cations in hole-doped \lsco or Ce$^{4+}$ cations substitutes R$^{3+}$ cations in electron-doped \rcco (R= La, Nd, Pr, Sm, Eu) \cite{Armitage2010}. Besides the Ce-doped concentration, superconductivity of the electron-doped cuprates depends heavily on the oxygen content. In 1995, $T_c$ of the under-doped \pcco (PCCO) by the protected annealing was enhanced comparable to that of the optimally doped one \cite{Brinkmann1995}. In 2008, the superconductivity in parent thin films $T^{\prime}$-R$_{2}$CuO$_{4}$ (R= Nd, Pr, Sm, Eu, Gd) was achieved by low-temperature annealing under the high-vacuum environment \cite{Matsumoto2008}. This momentous discovery seemingly challenges the commonly accepted model that high-$T_c$ superconductivity results from doping extra carriers into Mott insulators.

However, the intrinsic physics of superconductivity in the parent compounds is still under debate. Optical conductivity measurements on \pcod (PCO) thin films have disclosed that the AFM-correlated ``pseudogap" does not exist in this system, which indicates that the $T^\prime$-phase parent cuprates are metallic and superconductivity occurs at low temperature \cite{Chanda2014}. In fact, it is impossible to avoid oxygen defects during the sample preparation, which can introduce extra carriers into CuO$_2$ planes and lead to superconductivity. Wei \etal \cite{Wei2016} found that the superconducting $T^{\prime}$-La$_{2}$CuO$_{4}$ thin film can be tuned into an insulator by substituting Sr$^{2+}$ for La$^{3+}$. They suggest that $T^\prime$-phase parent cuprates are Mott insulators, while the intrinsic defects, most likely oxygen vacancies, are the sources of the effective carriers in these materials. Moreover, dynamical mean field theory (DMFT) studies demonstrate that the parent compounds may be described as weakly correlated Slater insulators rather than the strongly correlated charge transfer insulators \cite{Weber2010, Weber2010S}. Electrical transport measurement is a common but powerful method to explore the intrinsic electronic state in various superconductors. Plenty of studies in \rcco materials \cite{Jin2009, Jin2011, Dagan2005, Dagan2004, Zhang2016} support that the negative magnetoresistance (MR) and nonlinear magnetic field dependence of Hall resistivity are associated with antiferromagnetic order, Fermi surface reconstruction, quantum phase transition etc., which take a significant role in understanding the mechanism of high-$T_c$ superconductivity. Nevertheless, the systematic electrical transport investigations have been seldom addressed in parent system.

In this paper, we carried out systematical MR and Hall resistivity measurements on the superconducting PCO thin films which were synthesized on SrTiO$_3$ (STO) substrates by polymer assisted deposition method. With the decrease of oxygen content, the low-temperature MR gradually changes from negative to positive. Meanwhile, the low-temperature Hall coefficient $R_H$ undergo a  sign reversal, suggesting that dominant carriers vary from electrons to holes. All the behaviors are quite similar to those in \rcco materials. Furthermore, we plotted the $T_c$ versus $R_H$ for both PCCO and PCO samples, and found that their evolution could coincide with each other. Our results imply that the carriers induced by the oxygen removal benefit the superconductivity in $T^\prime$-parent cuprates.

\section{PREPARATION AND MEASUREMENTS}

Till now, the superconducting parent compounds have been successfully synthesized by metal-organic decomposition (MOD) \cite{Matsumoto2008} and molecular beam epitaxy (MBE) methods \cite{Yamamoto2010,Yamamoto2011,Wei2016}. In general, high-purity naphthenates used in MOD are extremely rare and expensive. Alternatively, MBE facility is very expensive, and it is of low-efficiency during the thin film preparation. Therefore, to develop a high-efficiency and inexpensive method is necessary for preparing the superconducting parent thin films. Polymer assisted deposition (PAD) is a newly developed method in recent years, which has the advantages of easy operation, low cost and stable precursor solutions \cite{Jia2004,Burrell2008}. Fig. 1(a) is a schematic diagram of PAD method. First, metal ions are electrostatically bound to the polymer in deionized water, forming a uniform and stable solution. Second, the solution is applied onto a substrate through either spin-coating or dipping. Finally the coated substrate is treated at a desired temperature in an oxygen environment to remove the polymer and enable the growth of the metal-oxide film.

\begin{figure}[ht!]
\includegraphics[width=1\columnwidth]{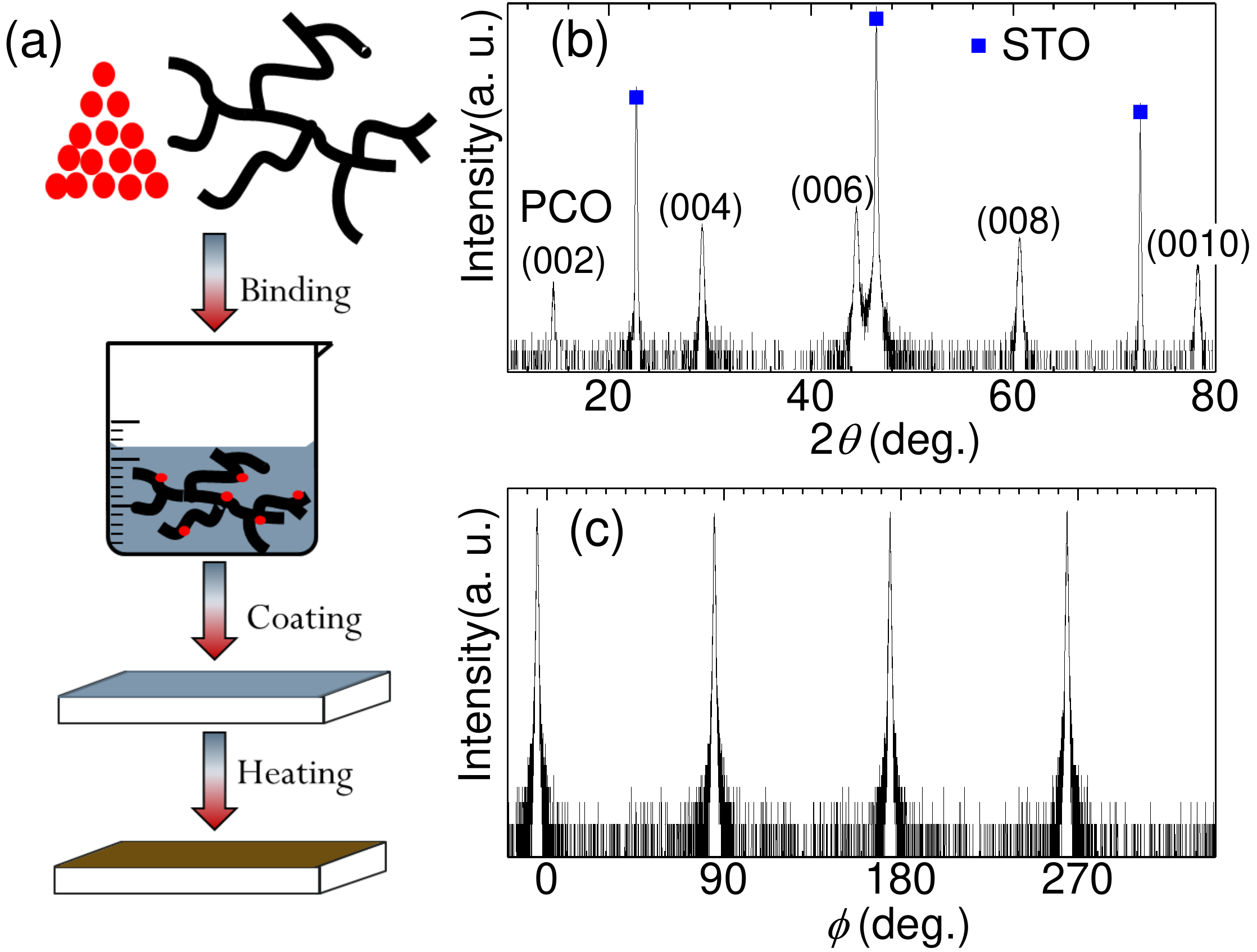}
\caption{(Color online) (a) The schematic diagram of polymer assisted deposition method. (b), (c) XRD $2\theta$-scans and $\phi$-scan of the \pcod film on the SrTiO$_3$ (001) substrate.} \label{fig:sample}
\end{figure}

To prepare high-quality superconducting PCO thin films, the metal ion sources are Pr and Cu nitrates, and the organic compounds are polyethylenimine (PEI) and ethylenediaminetetraacetic acid (EDTA). The concentration of Pr (Cu) metal ions in the solution was determined by inductively coupled plasma-atomic emission spectroscopy measurement. Then the precursor solution was obtained by mixing Pr solution and Cu solution at a certain stoichiometric ratio. In order to prevent the Cu absence, we made the precursor solution with 50\% Cu overdose. The precursor solution was spin coated on STO substrates. The polymer was pyrolyzed by heating gradually from the room temperature to 550 $\textordmasculine$C in the air. After that samples were sintered and crystallized at 850 $\textordmasculine$C for one hour in a tubular furnace under the oxygen pressure about 200 Pa. Finally the films were annealed in low-oxygen pressure about 1 Pa at various temperatures.

\begin{figure}[ht!]
\includegraphics[width=0.8\columnwidth]{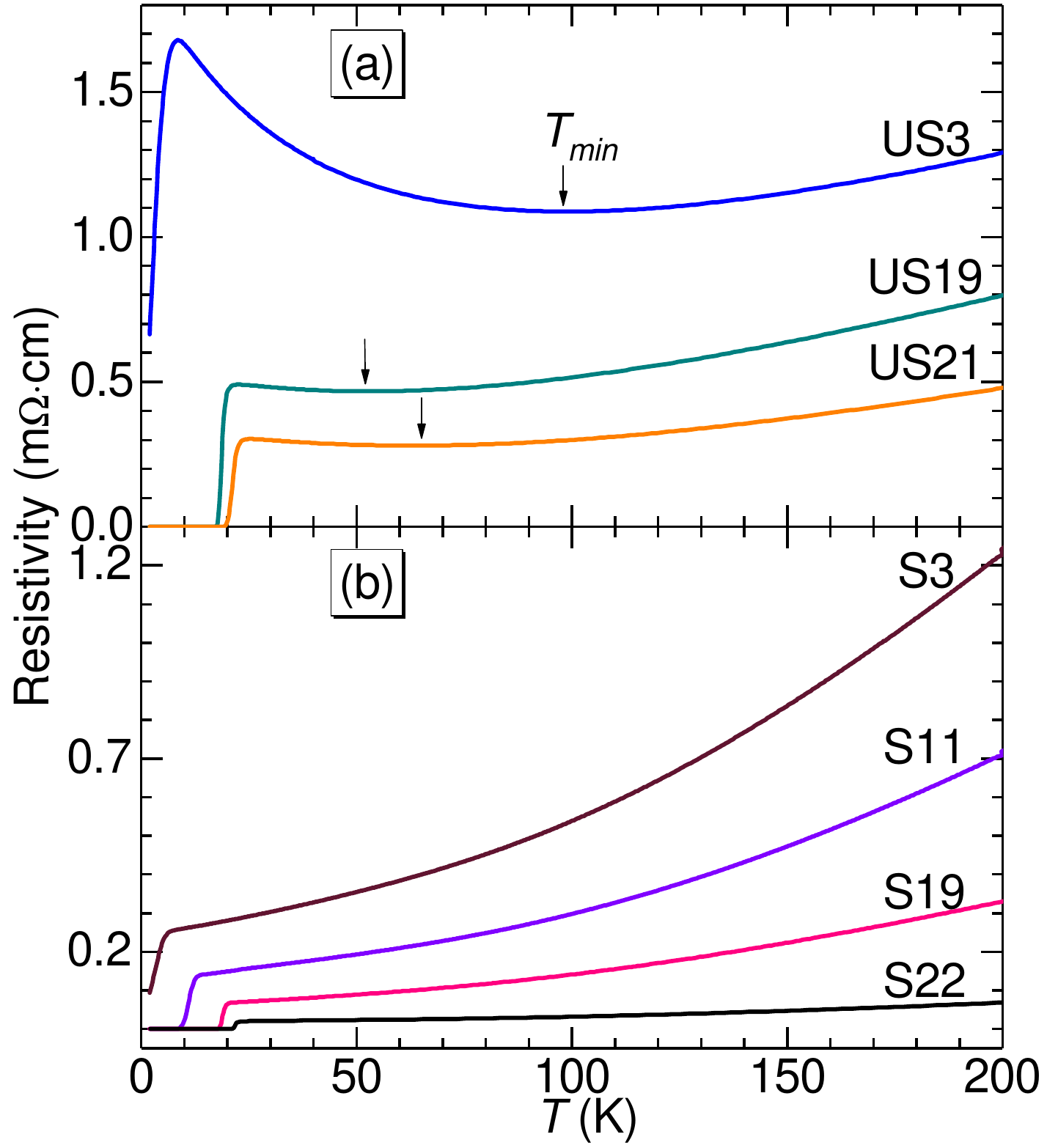}
\caption{(Color online) Resistivity as a function of temperature for \pcod films with various oxygen content. (a) The resistivity of US3, US19 and US21 shows
 an upturn at the temperature defined as $T_{min}$. (b) Resistivity versus temperature of S22, S19, S11 and S3 samples shows a good metallic behavior. The number in the sample name presents the superconducting transition temperature.} \label{fig:sample}
\end{figure}

To characterize the structure and crystalline of PCO thin films, we measured the $\theta$/$2\theta$ and $\phi$-scan of the samples using x-ray diffractometer. The XRD $2\theta$-scan of the PCO film on STO (001) is shown in Fig. 1(b). All peaks are sharp and can be indexed to (00l) of the $T^\prime$-structure, indicating that the film shows a single phase and c-axis orientation. Fig. 1(c) exhibits $\phi$-scan data with the approximately equal peak intensities, meaning that sample has a good epitaxy. Subsequently, electrical transport properties were measured by a 9 T physical property measurement system. In order to improve the measurement accuracy, all samples were patterned into standard Hall bridge. The range of the magnetic field is -9$\thicksim$9 T which was perpendicular to the film surface during the MR and Hall resistivity measurements.

\begin{table}[htbp]
\caption{\label{tab:test}Summary of the transport properties  and lattice structure constant $c_0$ of PCO films. $T_c$ is the superconducting transition temperature where the resistance becomes a half of the normal state value. $T_0$ represents the transition temperature of MR from negative to positive of US3, US21 and US19 samples. $RRR$ is the resistance ratio of $R$(300 K)/$R$(30 K). $c_0$ is the c-axis lattice constant.}
\setlength{\tabcolsep}{2.2mm}{
\begin{tabular}{cccccc}
\hline\hline
 Number & $T_c$(K) & $T_{min}$(K) & $T_0$(K) & $c_0$({\AA}) & $RRR$ \\
\hline
 US3 & $2.5$ & $98$ & $105\pm5$ & $12.213$ & $1.2$ \\
 US21 & $21.2$ & $65$ & $75\pm5$ & $12.212$ & $2.5$ \\
 US19 & $18.7$ & $52$ & $55\pm5$ & $12.211$ & $2.4$ \\
 S22 & $21.9$ & $$ & $$ & $12.205$ & $4.8$ \\
 S19 & $18.9$ & $$ & $$ & $12.202$ & $7.8$ \\
 S11 & $11.0$ & $$ & $$ & $12.201$ & $8.0$ \\
 S3 & $2.7$ & $$ & $$ & $12.199$ & $7.2$ \\
\hline\hline
\end{tabular}}
\end{table}

\section{RESULTS AND DISCUSSIONS}

In Fig. 2, we present the temperature dependence of the resistivity for various PCO thin films. With the decrease of the temperature, resistivity for US3, US21 and US19 samples shows an upturn at the temperature  $T_{min}$. Considering that $c_0$ should have a positive correlation with the relative oxygen content of the samples \cite{Matsumoto2009}, we calculated the c-axis lattice constants $c_0$ of PCO films (see Table I) by Bragg diffraction formula from the XRD data. It is worth mentioning that $T_{min}$ gradually decrease as the $c_0$ reduces and the upturn behavior disappears for S22, S19, S11 and S3.

\begin{figure}[ht!]
\includegraphics[width=1\columnwidth]{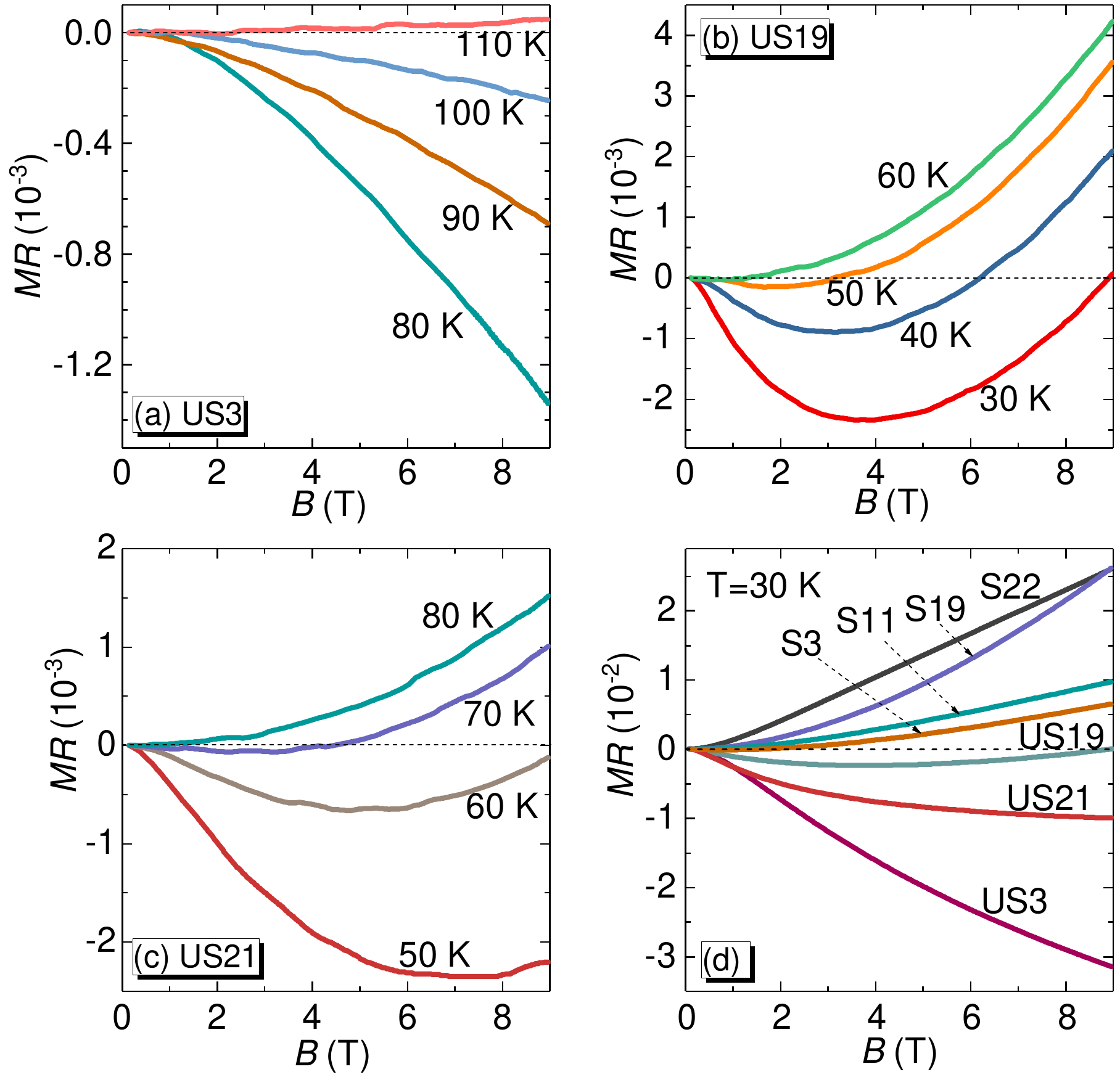}
\caption{(Color online) (a), (b), (c) MR versus magnetic field $B$ for US3, US19 and US21, respectively. (d) MR curves for the various \pcod films at $T$=30 K.} \label{fig:sample}
\end{figure}

In the underdoped region of Ce-doped cuprates, the low-temperature resistivity also shows an upturn which gradually disappears with the increase of Ce concentration \cite{Dagan2005, Finkelman2010}. High magnetic field can suppress this upturn behavior, resulting in a negative MR \cite{Zhang2016}. Fig. 3 displays the field dependence of MR, with applied field parallel to the c-axis $(B//c)$. The MRs of US3, US21 and US19 samples are negative at the low temperature and turn into positive at the high temperature as shown in Fig.3 (a), (b) and (c). We define the transition temperature as $T_0$ which performs a consistent tendency with $T_{min}$ (as shown in Table I). Besides, the low-temperature negative MR fades away as reducing the oxygen content(Fig. 3(d)). We conjecture that the mechanisms of the upturn and negative MR in US3, US21 and US19 samples are the same as those in the under-doped region of Ce-doped cuprates.

The Hall coefficient $R_H$ can illuminate the nature of material carriers. For example, the low-temperature $R_H$ in the \rcco system varies from negative to positive as the Ce-doped concentration increases, corresponding to the carriers varying from electrons to holes \cite{Dagan2004, Charpentier2010, Jin2009}. Fig. 4(a) exhibits $R_H$ as a function of temperature for various samples. The $R_H$s of US3, US19 and US21 are negative in the whole temperature range, suggesting that the electron is predominant in the transport process. The sign reversal of $R_H$ arises in S22 and S19 as the temperature increases, i.e. from positive to negative, which reveals a competition between electron and hole carriers. The $R_H$s of S11 and S3 are positive below 150 K, indicating that the dominant carriers are holes.

\begin{figure}[ht!]
\includegraphics[width=1\columnwidth]{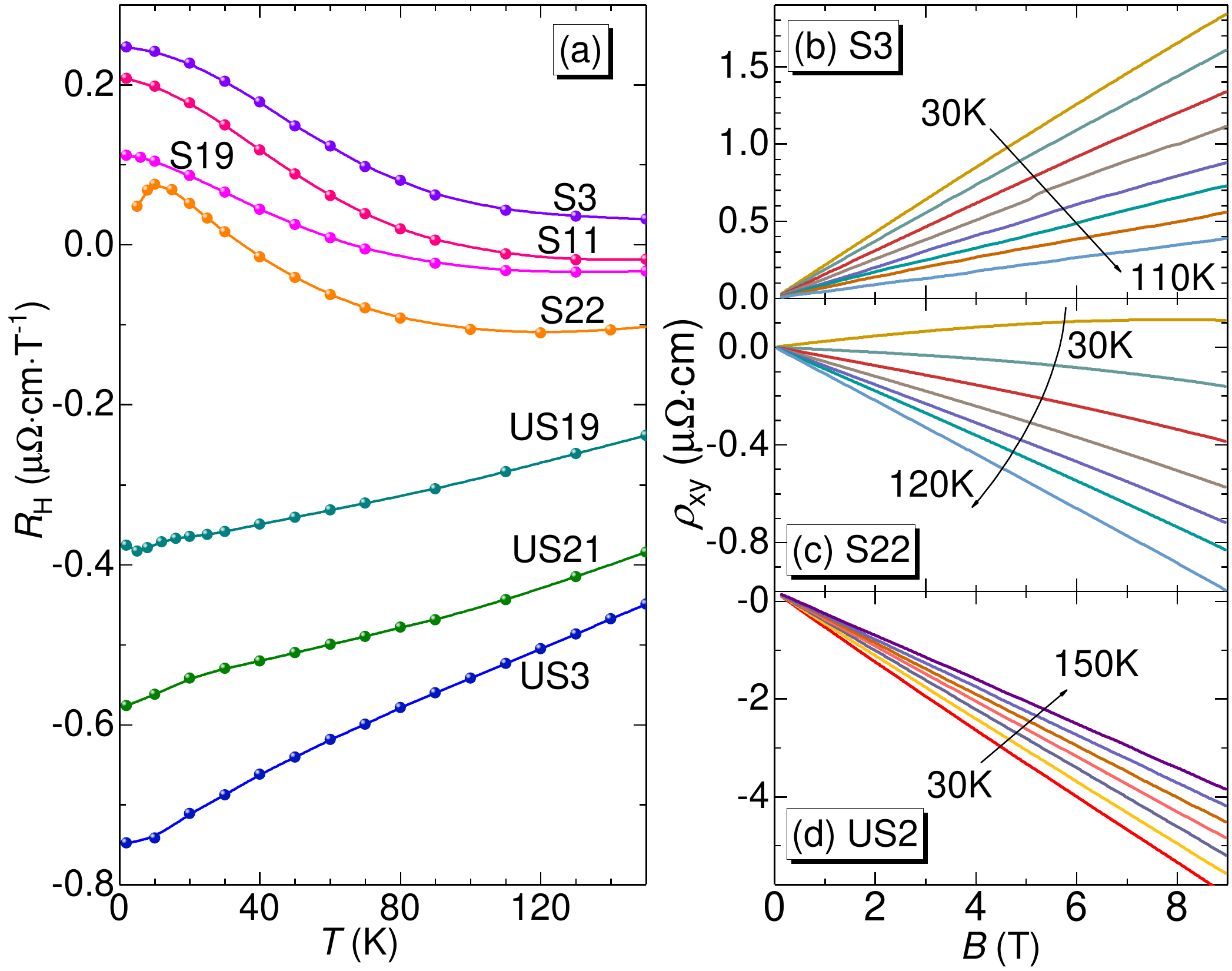}
\caption{(Color online) (a) Normal state Hall coefficient $R_H$ versus temperature for \pcod films measured at $B$=9 T (B//c). (b), (c), (d) Hall resistivity $\rho_{xy}$ versus $B$ for S3, S22, US3.} \label{fig:sample}
\end{figure}

Fig. 4(b), (c) and (d) display the Hall resistivity $\rho_{xy}$ at various temperature for US3, S22 and S3, respectively. For US3 and S3, the $\rho_{xy}$ as a function of magnetic field exhibits a good linear behavior, indicating that the samples have a single band characteristic. The low-temperature $\rho_{xy}$ of S22 sample is nonlinear to the magnetic field, which implies the coexistence of electron and hole band. For Ce-doped cuprates, Fermi surface of the optimum doped samples is reconstructed by antiferromagnetic order \cite{Dagan2004, Lin2005, Sarkar2017}. Therefore, we can surmise that Fermi surface of the parent compounds can be gradually adjusted by changing the oxygen content.

As mentioned above, it can be concluded that, for MR and Hall resistivity, the effects of varying oxygen content in PCO thin films are similar to that of varying Ce concentration in PCCO. Such phenomenon indicates that the oxygen removal can induce carrier doping. This finding seemingly supports the scenario that superconductivity in parent compounds is obtained via doping a Mott insulator, symmetric to the hole-doped side. We note that this view has been clarified in the recent ARPES results as well \cite{Horio2018ARPES, Wei2016}.

Unfortunately, it is extremely difficult to measure the accurate oxygen content which also plays a non-ignorable role in \rcco. Song \etal \cite{song2017} studied \plcco samples with different Ce concentrations and annealing processes using ARPES. They found that effective electron number estimated from Fermi surface volume could unify the effect of Ce doping and deoxygenation. For electrical transports, $R_H$ is the most qualified quantity reflecting the information of carriers. Therefore, we try to set $R_H$ as the horizontal axis to study the evolution of $T_c$ in both PCCO and PCO systems. As shown in Fig. 5, we find that all the data from distinct groups, various measuring conditions and different samples obey the same rule, i.e. $T_c$ increases at first and then decreases as $R_H$ increases. Especially, $T_c$ always reaches its maximum near $R_H$= 0, which indicates that the balance between the electron and hole bands is beneficial to the superconductivity of electron-doped cuprates, consistent with the electrostatic tuning results of Pr$_{1.85}$Ce$_{0.15}$CuO$_{4}$ \cite{Jin2016}.

\begin{figure}[ht!]
\includegraphics[width=0.8\columnwidth]{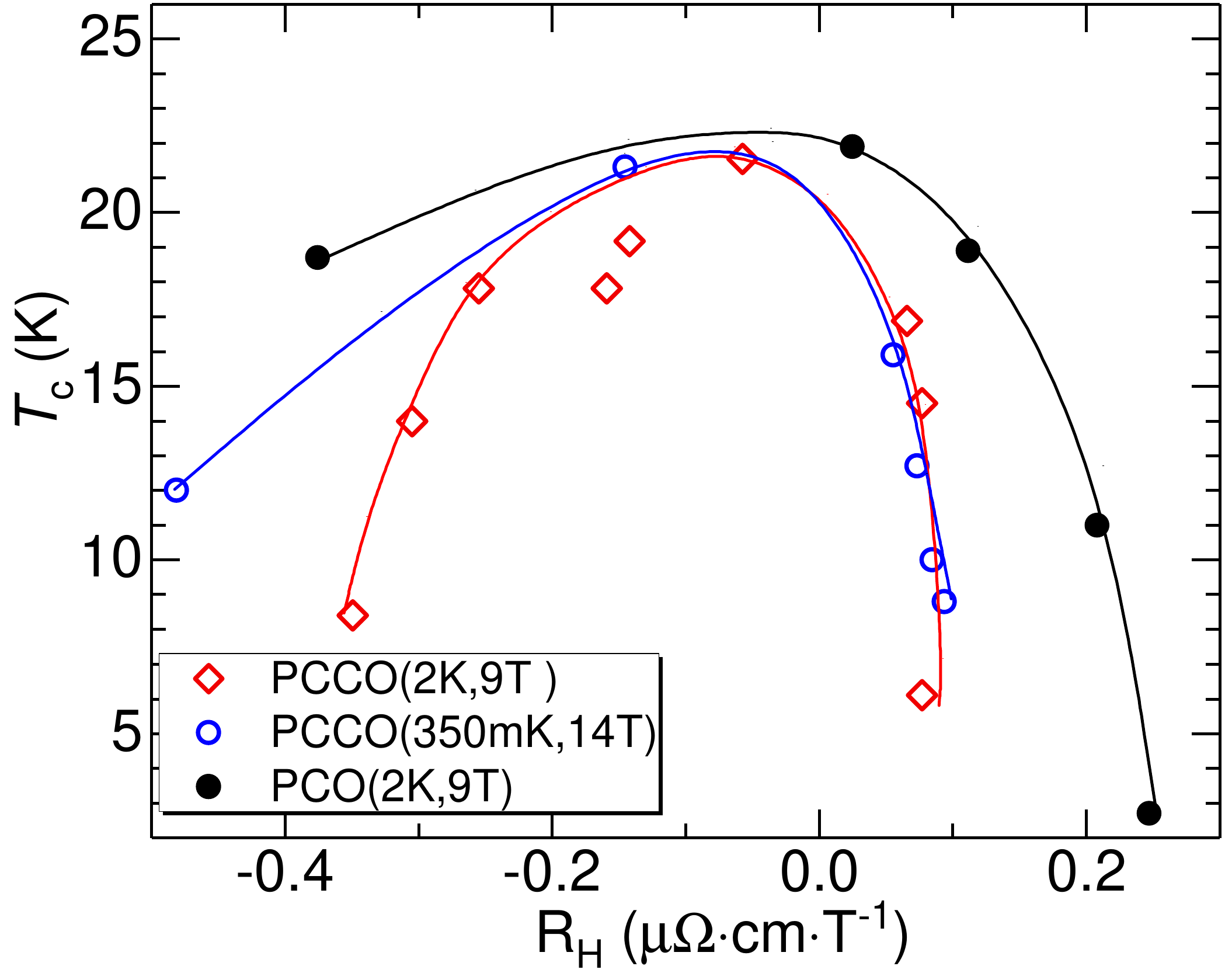}
\caption{(Color online) $T_c$ versus $R_H$ of both PCO and PCCO. PCCO (2 K, 9 T) and PCCO (340 mK, 14 T) data derive from ref. 12 and ref. 20 respectively. The solid curve is a guide to the eye.} \label{fig:sample}
\end{figure}

For the underdoped PCCO samples, the $R_H$ rapidly decreases as temperature decreases, independent of the magnetic field \cite{Li2007,Dagan2004}. Therefore, the clear difference between two group data of PCCO is mainly attributed to the temperature distinction. Of obviously, we cannot exclude the influence of antiferromagnetic order and disorder. The PCCO data show a perfect overlap in the overdoped region because the $R_H$ is hardly dependent on the temperature and linear with magnetic filed in this region. As shown in Fig. 5, we can see that the $T_c$ of PCO is always greater than that of PCCO samples at a fixed $R_H$ value. This can be explained as the following. In PCO and PCCO, the change in $T_c$ is mainly due to the change of carrier concentration and structural disorder i.e.
$\Delta$$T_c$=$\Delta$$T_c$($R_H$) + $\Delta$$T_c$(disorder).
We assume that the $\Delta$$T_c$($R_H$) induced by Ce doping or oxygen changes is identical for both PCO and PCCO samples. The disorder in PCCO is caused by Ce doping and O content changes. In contrast, the disorder in PCO can be just attributed to changing the oxygen content. So the absolute $\Delta$$T_c$(disorder) of PCO is less than that of PCCO at a fixed $R_H$ value. This suggests the disorder introduced by Ce doping is more than that by deoxygenation. It can be understood by that Ce doping causes a structure distortion and brings more disorder in PCCO. This finding can explain why $T_c$ in the parent  compounds are even higher than that of the optimal Ce-doped samples \cite{Matsumoto2009} and also support that the superconductivity in $T^\prime$-phase parent cuprates stems from doping effect induced by the oxygen removal.

In summary, a series of high-quality superconducting PCO thin films have been prepared by polymer assisted deposition method. The magnetoresistance and Hall effect of superconducting PCO thin films with various oxygen content are systematically studied. It is found that for the magnetoresistance and Hall resistivity, the effect of oxygen removal in PCO is similar to that of the Ce doping in \rcco. In addition, we find that Hall coefficient $R_H$ is an efficient parameter to depict the evolution of $T_c$ in the electron-doped cuprates. We argue that doped electrons induced by the oxygen removal are the cause of the superconductivity in $T^\prime$-parent compounds.

\section{ACKNOWLEDGMENTS}

This work was supported by the National Key Basic Research Program of China (Grants No. 2015CB921000, No. 2016YFA0300301, No. 2017YFA0303003 and No. 2018YFB0704100), the National Natural Science Foundation of China (Grants No. 11674374 and No. 11474338), the Key Research Program of Frontier Sciences, CAS (Grant No. QYZDB-SSW-SLH008) and the Strategic Priority Research Program of the CAS (Grants No. XDB07020100 and No. XDB07030200), the Beijing Municipal Science and Technology Project (Grant No. Z161100002116011).

\nocite{*}
\bibliography{refpco}

\end{document}